\documentclass[cameraready]{vgtc}             % electronic version

\ifpdf%                                % if we use pdflatex
\pdfoutput=1\relax                   % create PDFs from pdfLaTeX
\usepackage{graphicx}                % allow us to embed graphics files
\DeclareGraphicsExtensions{.pdf,.png,.jpg,.jpeg} % for pdflatex we expect .pdf, .png, or .jpg files
\else%                                 % else we use pure latex
\usepackage{graphicx}                % allow us to embed graphics files
\DeclareGraphicsExtensions{.eps}     % for pure latex we expect eps files
\fi%

\graphicspath{{figures/}{pictures/}{images/}{./}} % where to search for the images

\usepackage{microtype}                 % use micro-typography (slightly more compact, better to read)
\PassOptionsToPackage{warn}{textcomp}  % to address font issues with \textrightarrow
\usepackage{textcomp}                  % use better special symbols
\usepackage{mathptmx}                  % use matching math font
\usepackage{times}                     % we use Times as the main font
\usepackage{cite}                      % needed to automatically sort the references
\usepackage{tabu}                      % only used for the table example
\usepackage{booktabs}                  % only used for the table example
\usepackage{multirow}
\usepackage{xcolor}
\usepackage{listings}

\lstdefinestyle{yaml}{
     basicstyle=\color{blue}\footnotesize,
     rulecolor=\color{black},
     string=[s]{'}{'},
     stringstyle=\color{blue},
     comment=[l]{:},
     commentstyle=\color{black},
     morecomment=[l]{-}
 }

\usepackage{xspace}
\usepackage[colorlinks=true, linkcolor=blue, urlcolor=blue, citecolor=blue]{hyperref}

\onlineid{0}

\vgtccategory{Research}

\newcommand{\name}{\texttt{TOBY}\xspace}
\newcommand{\xaip}{\texttt{XAIP}\xspace}
\newcommand{\vamhri}{\texttt{VAM-HRI}\xspace}
\newcommand{\macq}{\texttt{MACQ}\xspace}

\title{\name: A Tool for Exploring Data in Academic Survey Papers}

\author{Tathagata Chakraborti, Jungkoo Kang\\ %
\scriptsize IBM Research AI %
\and Christian Muise\\ %
 \scriptsize Queen's University %
\and Sarath Sreedharan\\ %
 \scriptsize Colorado State University\\[-1ex]%
\and Michael Walker, Daniel Szafir\\ %
 \scriptsize University of North Carolina at Chapel Hill%
\and Tom Williams\\ %
 \scriptsize Colorado School of Mines %
}

\abstract{
This paper describes \name, a visualization tool that helps an user
explore the contents of an academic survey paper. The visualization
consists of four components: a hierarchical view of taxonomic data in the survey,
a document similarity view in the space of taxonomic classes, a network view of 
citations, and a new paper recommendation tool.
In this paper, we will discuss these features in the context of three separate
deployments of the tool.
} % end of abstract

\begin{document}
\maketitle


\section{Introduction}
\label{sec:intro}

\maketitle

In her presidential address \cite{gil2022will} at AAAI
(Association for Advancement of Artificial Intelligence) 2020, 
one of the premier conferences in the field of artificial intelligence (AI),
then-president Yolanda Gil asked: {\em ``Will AI write scientific papers in the future?''}.
The question was posed to facilitate 
an exploration of the influence that AI algorithms, 
from process management to knowledge discovery, increasingly 
have on our scientific endeavours. 
The intent of the question is not that AI algorithms will be 
writing their own papers soon but rather 
to acknowledge an emerging theme of {\em collaboration between AI and the 
scientist} in terms of making the process of scientific discovery easier,
faster, and more efficient.

One of the most powerful emerging uses of AI within the scientific process itself
is to facilitate the exploration of the existing body of work in a field,
in ways that stimulate further research and familiarize the researcher 
with the state of the art in their field.
For example, preliminary visualization tools have begun to emerge that enable 
researchers 
to visualize and explore the historical record of work published
at longstanding conferences such as 
NeurIPS (Neural Information Processing Systems) \cite{strobeltinteractive, strobeltinteractive-new}, or within specific subfields such as constraint programming
and satisfiability theory \cite{kotthoff2018quantifying}.
Outside of specific conference, there is a long history of work
on the topic of visualizing scientific literature 
in general \cite{federico2016survey, zhang2018survey},
including popular tools such as SurVis, PaperQuest, 
and others \cite{beck2015visual, ponsard2016paperquest, dattolo2022authoring}.

We explore a new instance of this theme. We
use data from a survey paper and generate different forms of visualizations 
to explore it but with a particular twist: the ability
to identify gaps in literature based on the provided data. 
The tool we present to facilitate this task, \name 
provides multiple visualizations that offer complementary views into the survey 
data: classification hierarchies, document similarities, and influence networks.
Most excitingly, 
with the help of a SAT-encoding \cite{kissat} of the survey data,
the tool automatically surfaces papers yet unwritten, 
to drive the researcher forward into new areas of research.
We thus make two contributions:

\begin{figure*}
\centering
\includegraphics[width=\linewidth]{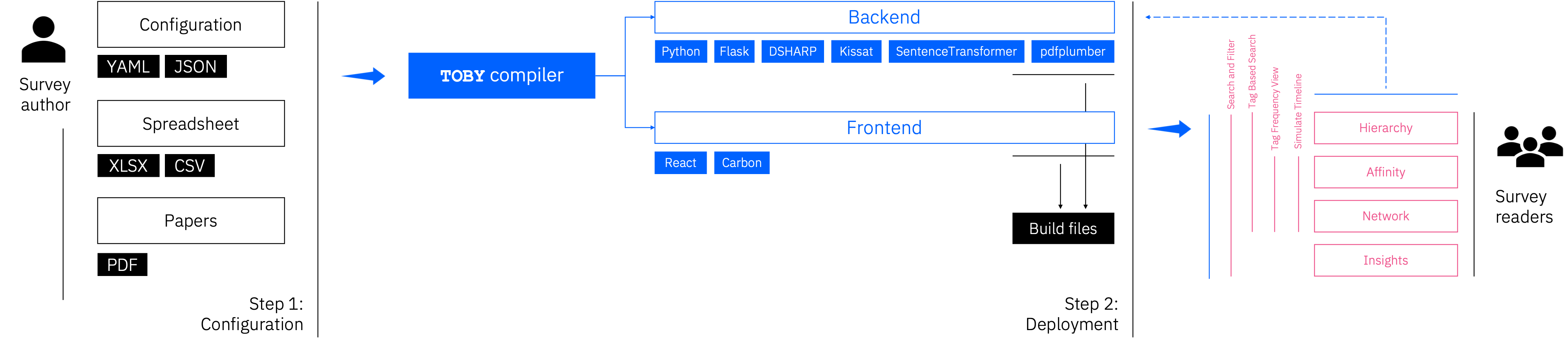}
\caption{\name technology stack and flow of control.}
\label{fig:toby}
\end{figure*}

\begin{figure*}[tbp!]
\includegraphics[width=\linewidth]{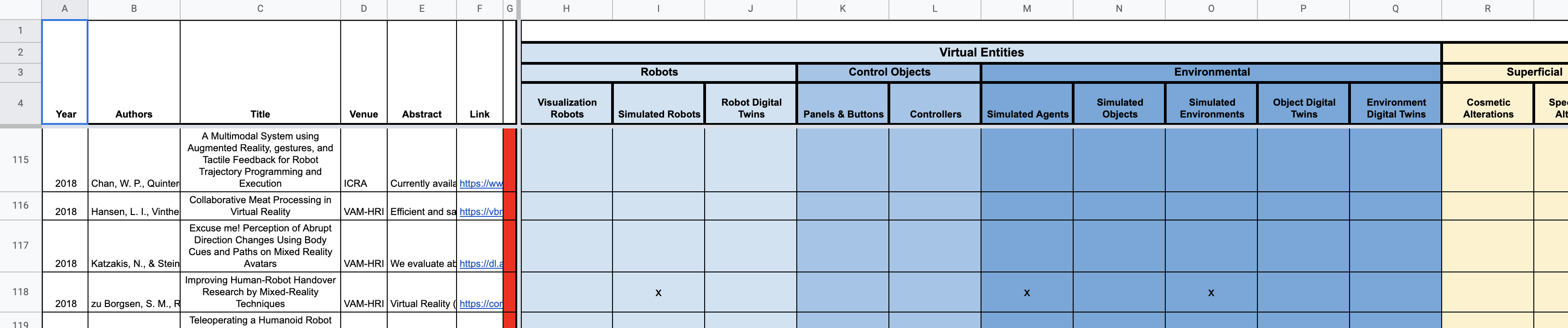}
\caption{
A spreadsheet capturing survey data in terms 
of a hierarchical classification of papers on a topic.
From the deployment of: \cite{walker2022virtual}.
}
\label{fig:spreadsheet}
\end{figure*}

\begin{itemize}
\item[1.]
The primary contribution (Section \ref{sec:toby}) of this work is a tool (\name)
that enables exploration of a encoded literature survey.
To facilitate this exploration, \name 
leverages automation and AI tools to extract data from
PDFs, generate document embeddings, and identify gaps in coverage of topics.

\item[2.] 
The secondary contribution (Section \ref{sec:insights}) of this work is a
built-in SAT-encoding 
that allows \name to help users identify papers yet to be written according to a survey.
To the best of our knowledge, this is the only tool, among software of its class \cite{RushStrobelt2020, strobeltinteractive-new, strobeltinteractive, kotthoff2018quantifying, beck2015visual, ponsard2016paperquest}, that has this capability.
\end{itemize}

\subsection{Target Audience and Use Cases}
\label{subsec:audience}

Literature surveys are critical to scientific researchers, as they enable
communities to onboard new scholars, and help existing researchers monitor
the state of the art in their areas of expertise.
The objective of our tool is help in that process.
Among researchers in general, an important 
subclass of use cases for us is to empower new students to quickly understand
the state of the art on a topic as they try to come up with projects
for their classes and their research. 

While the previous two use cases focused on the reader of a survey as the user,
the tool can also help the authors of a survey to better present their work
and reach their audience. 
Finally, similar to the uses 
in \cite{strobeltinteractive, strobeltinteractive-new, RushStrobelt2020}
and not confined to survey data only,
we envisage that \name can be used by chairs of conferences to not only visualize 
data at their event but also be empowered to make data-driven decisions 
during the review and acceptance stage to identify how would-be papers in
their conference
will represent (or fail to cover) the diversity of topics in their field.

\subsection{Current Deployments}
\label{subsec:deployments}

\name has already been used for deployments
on three different surveys.
The rest of the paper will use data and illustrations,
with permission, 
from each of these deployments.

\vspace{10pt}
\noindent {\bf \vamhri} Virtual, Augmented, and Mixed-Reality 
for Human-Robot Interaction \cite{walker2022virtual}: 
This is a rapidly emerging sub-area of interest in the field of 
human-robot interaction, 
as improving mixed-reality technologies enable hitherto unexplored \cite{williams-cube} 
methods of interactions between humans and robots. 
This survey and taxonomy analyzes the journey of the VAM-HRI community at 
HRI, founded in 2017 \cite{williams-ai-mag}, and beyond.
Link: \textcolor{blue}{\href{http://vamhri.com}{vamhri.com}}.

\vspace{10pt}
\noindent {\bf \macq}
Model Acquisition for Planning \cite{keps22}:
This survey looks at the field of model learning for AI planning algorithms -- a topic intriguingly placed
at the intersection of reinforcement learning and automated planning. 
The unique nature of this topic, as a mixture of learning and modeling, makes this a primary target for 
our tool due to obvious blind spots of researchers coming from either area of expertise. 
This deployment accompanies the recently open-sourced tool \macq: \url{https://github.com/QuMuLab/macq}, 
aimed to be the one-stop shop for data-driven model acquisition for automated planning domains.
Link: \textcolor{blue}{\href{https://macq.planning.domains}{macq.planning.domains}}.

\vspace{10pt}
\noindent {\bf \xaip}
Explainable AI Planning \cite{chakraborti2019explicability, chakraborti2020emerging}:
The world of Explainable AI is rapidly expanding in scope from classification tasks 
to more complex decision-making processes where automated algorithms play an outsized role. 
The community engaged at the Explainable AI Workshop at the International Conference of Automated Planning 
and Scheduling, the premier conference on planning and sequential decision-making algorithms, 
has been at the forefront of this evolution \cite{hoffmann2019explainable, xaip}.
This deployment captures two
complementary threads of work in this area: explaining agent behavior
and designing agent behaviors that are inherently explainable
\cite{chakraborti2019explicability, chakraborti2020emerging}.
Link: \textcolor{blue}{\href{https://explainableplanning.com}{explainableplanning.com}}.

\section{System Overview}
\label{sec:toby}

Figure \ref{fig:toby} illustrates the technology stack and 
flow of control 
in \name. The \name compiler does most of the heavy lifting 
in generating all the required
artifacts automatically from the user inputs. 
We describe this process below, with examples
of the produced outputs from the surveys described in Section \ref{subsec:deployments}.

\subsection{Setting Up}
\label{subsec:start}

\noindent {\bf Survey Data:}  
\name requires the survey data in the form a spreadsheet where the data
is organized as a list of papers (rows) underneath a set of categories (columns) 
where for each row, each cell is marked depending on whether that paper belongs to the category on that column or not.
This is a pretty generic format (an example is presented in Figure \ref{fig:spreadsheet}).
The \name compiler will read hierarchies present in this spreadsheet automatically
based on how it is configured.

\vspace{5pt}
\noindent {\bf Configuration:}
The first part of the configuration deals with metadata of the survey paper being 
used, relevant links to read the paper, contribute to the data, and so on. 
This is followed by a set of directives for where to look for relevant
data in the spreadsheet: the rows and columns (for either of the taxonomy area 
or the paper list) indicate the start and end of 
where \name should read from in the spreadsheet, 
as well as rows / columns to ignore if required. 

For example, in Figure \ref{fig:config}, we are asking \name to read the ``Taxonomy'' hierarchy 
from rows 1-4 and columns 69-146, and the rows 7-184 (excluding 141 and 151) for the list of papers. 
The \texttt{key\_map} entry indicates where (columns) 
the paper metadata is documented (e.g. the title is in column 3).
There can be multiple taxonomies provided
% (e.g. Table \ref{tab:deployments}) 
-- the one marked as default will be used as the underlying classification scheme in the other views.

\vspace{5pt}
\noindent {\bf Documents:}  
Finally, the user can optionally provide a set of PDFs for the papers contained in their survey. 
These are used to automatically generate a citation network 
by parsing the bibliography sections of the paper provided
and fuzzy matching them to the metadata of the papers in the survey.

\begin{figure}[!ht]
\small
\begin{lstlisting}[style=yaml]
tab_name: Taxonomy
title_text: Taxonomy View of VAM-HRI Design Elements
input_file:
  filename: /path/to/spreadsheet.xlsx
  active_worksheet: main
papers_list:
  key_map:
    title: 3
    abstract:
    authors: 2
    venue: 1
    year: 0
  rows:
    start: 7
    stop: 184
    exclude:
      - 141
      - 151
taxonomy:
  rows:
    start: 1
    stop: 4
  columns:
    start: 69
    stop: 146
\end{lstlisting}
\caption{
YAML configuration of the Taxonomy tab. From the deployment 
of: \cite{walker2022virtual}.
}
\label{fig:config}
\end{figure}

\begin{figure*}[tbp!]
\includegraphics[width=\linewidth]{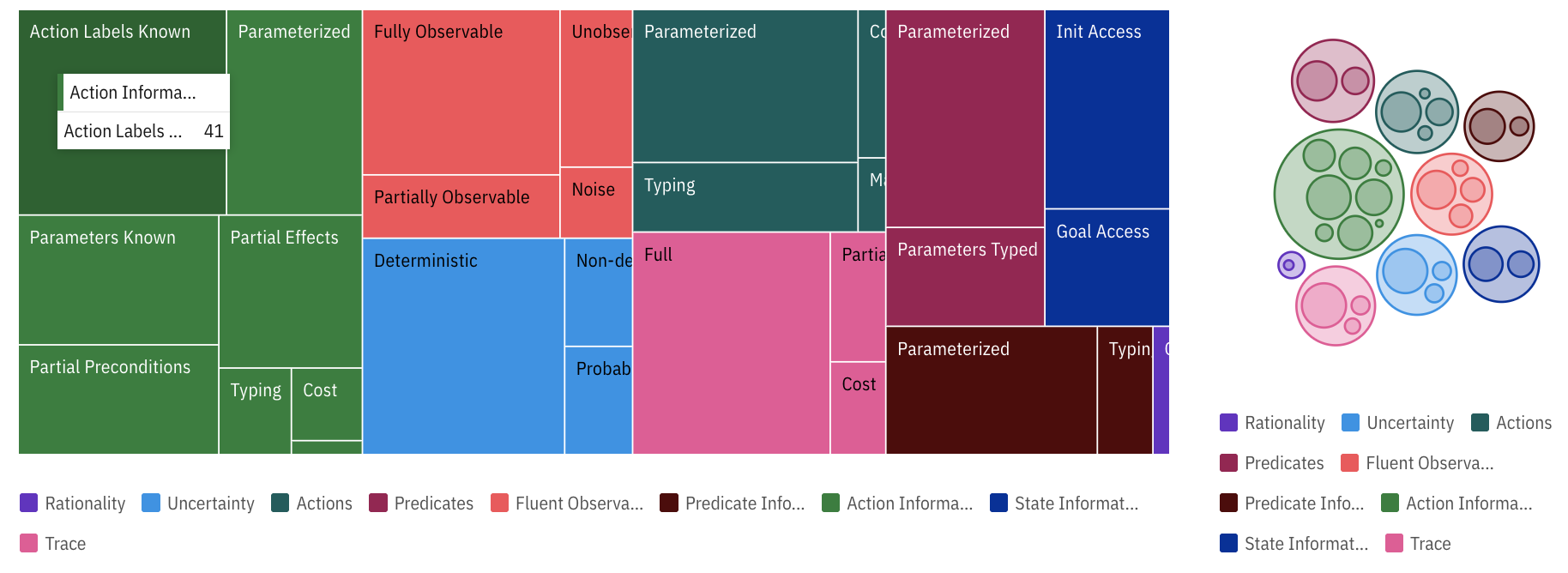}
\caption{
Treemap of a level of the taxonomy. 
From the deployment of \cite{keps22}.
}
\label{fig:treemap}
\end{figure*}

\begin{figure*}[tbp!]
\includegraphics[width=\linewidth]{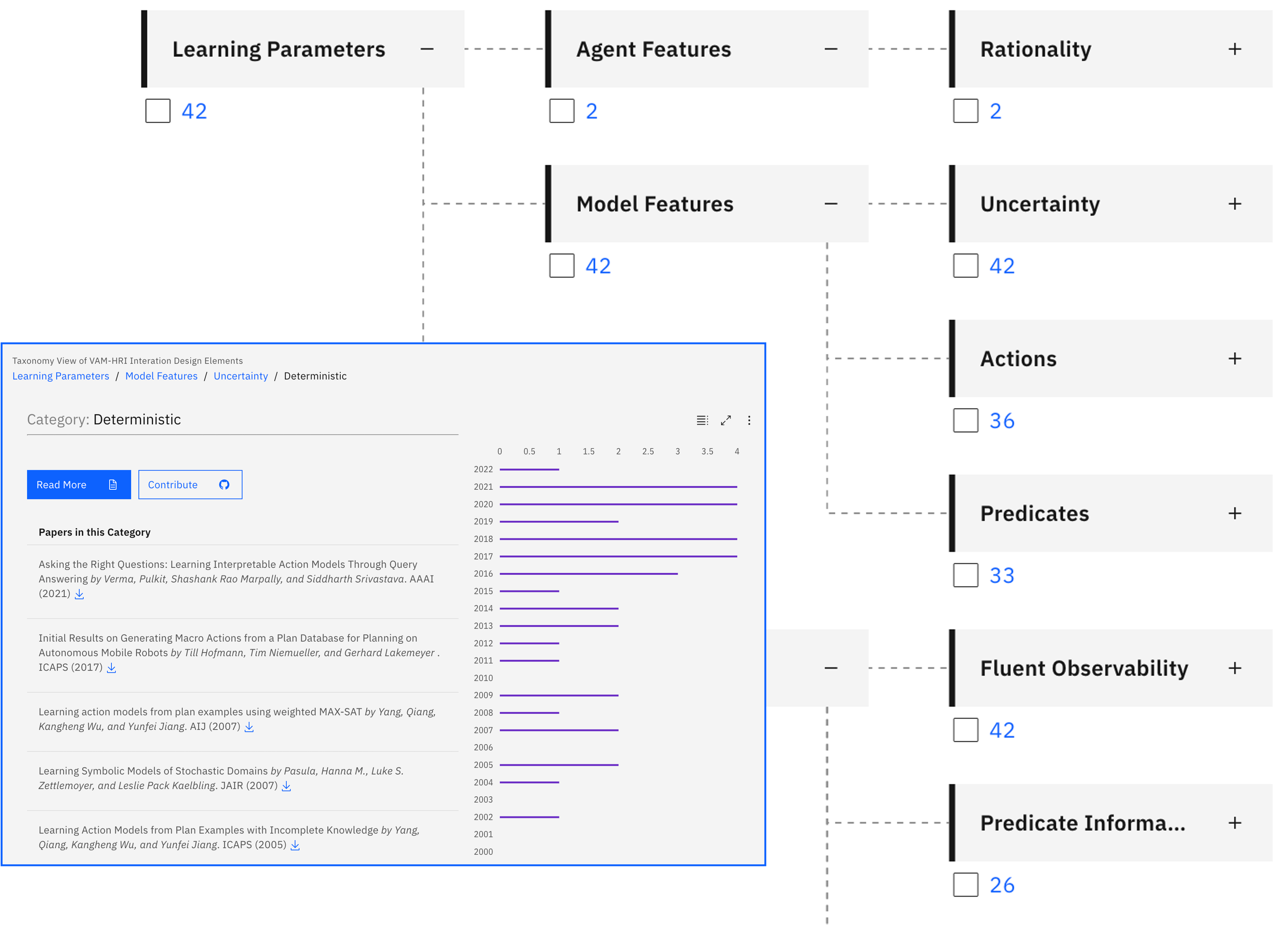}
\caption{
Hierarchical view of the taxonomic classes (inset showing timeline
of research of a particular class). 
From the deployment of \cite{keps22}.
}
\label{fig:hierarchy}
\end{figure*}

\begin{figure*}[tbp!]
\includegraphics[width=\linewidth]{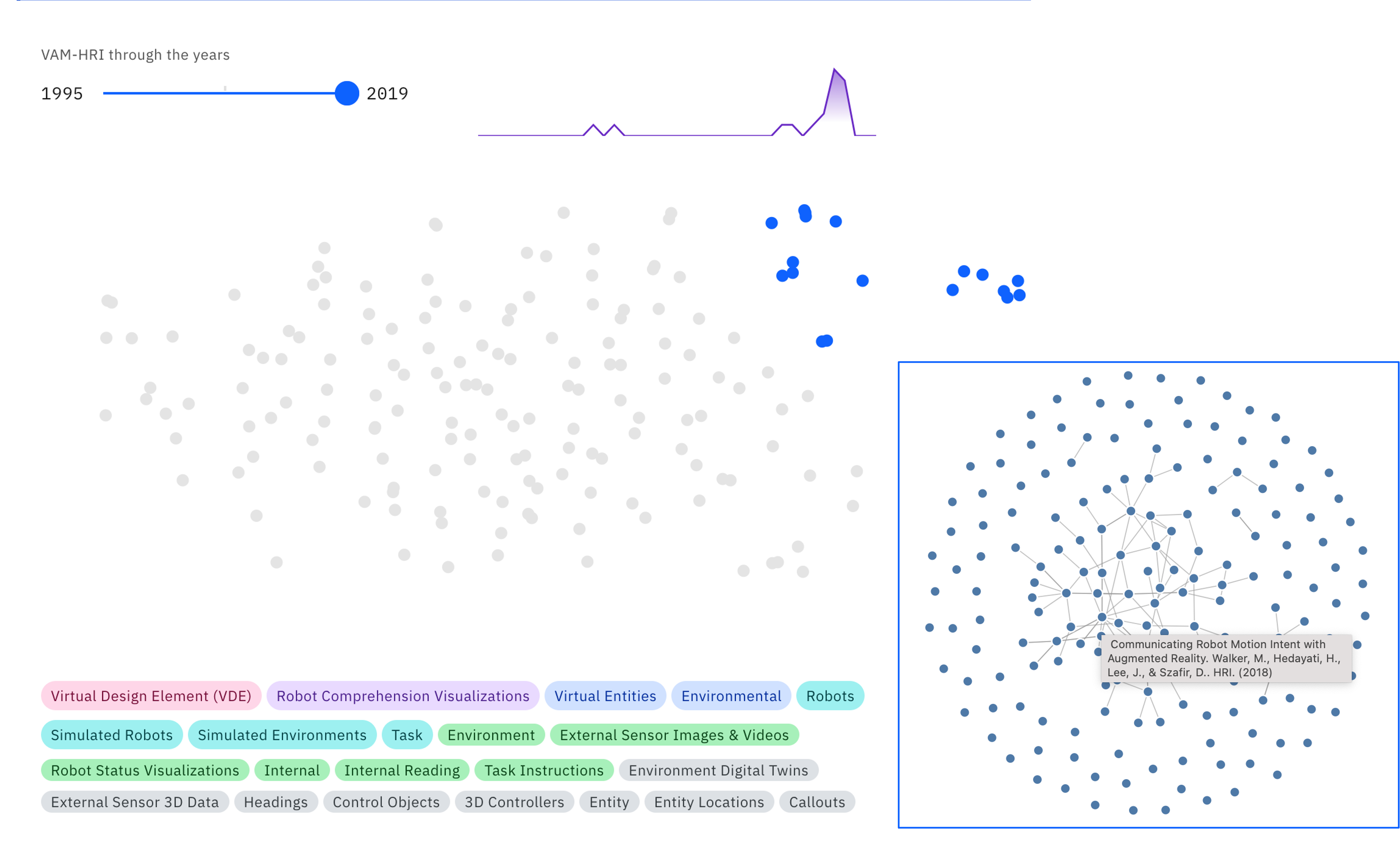}
\caption{
Affinity and network view.
From the deployment of \cite{walker2022virtual}.
}
\label{fig:dual}
\end{figure*}

\subsection{Four Different Views into the Data}
\label{subsec:views}

Some basic features span across one or more (or all) views in \name, as indicated in Figure \ref{fig:toby}.
This includes the ability to search papers with (conjunctions or disjunctions, as desired, of) keywords,
filtering results by range of years, selection of one or more taxonomic classes from the survey data, and so on.
An advanced cross-view feature is that of timeline simulation that allows the reader to 
simulate the evolution of the field across time.
This feature mimics the feature from the visualization in \cite{strobeltinteractive-new},
but extended in \name to cover the network view as well.
During this simulation, as the user drags their control over the timeline,
the visualization in taxonomic space (or graph in case of the citation
network) is re-rendered according to the papers contained in the timeline.
Additionally, a graph of the number of papers 
during the simulation are shown in the form
of a sparkline over the selected period of time.

\vspace{5pt}
\noindent {\bf Hierarchy View}
The primary view provides a taxonomic account 
of the various topics identified in the field and how
papers are classified along those topics -- this is 
shown in Figures \ref{fig:treemap} and \ref{fig:hierarchy}.

\vspace{5pt}
\noindent {\bf Affinity View} 
The next view displays the papers in a survey in the space of features. 
The document embedding is computed according to
the approach in \cite{specter2020cohan},
and adopted from a similar application in \cite{RushStrobelt2020}.
In this view, the user can select subsets of papers in
feature space and filter by features by clicking on the tags. 
Figure \ref{fig:dual} provides an illustration
of the user exploring a cluster of papers on the top right
which are summarized by frequency of tags to belong
to robot comprehension and virtual environment features
from \cite{walker2022virtual}.

\vspace{5pt}
\noindent {\bf Network View} 
From the PDF documents of the papers,
we also automatically extract, using \cite{pdfplumber}
(more details later in Section \ref{subsec:perf-network}), 
a citation network to illustrate the most influential hubs in the topic of a survey.
This is shown in Figure \ref{fig:dual} (inset) where the user
is hovering over an influential paper \cite{walker2018communicating} in the field
of \vamhri, automatically identified by \name 
from the documents
in \cite{walker2022virtual}.
Similar to the affinity view, here too the user 
can modify the network view using the feature space 
as well as simulate 
how the network evolves over time.

\vspace{5pt}
\noindent {\bf Insights View}
Finally, the insights view takes the reader through 
the most and least popular 
classes of papers in the survey data, 
as well as classes that have no papers yet. 
An advanced feature in this view is \name's ability to help 
its users identify papers, in terms of 
the taxonomic classes in the survey, that do not exist yet.
To this end, one or more new papers
are generated and presented as follows:

\begin{itemize}

\item[-] The first part of the exposition describes the newly generated paper(s) in terms of its features. 
The position of the new paper in the classification hierarchy from the survey is displayed.

\item[-] The hypothetical paper is now visualized in feature space: this view
shows where the new paper belongs when all the papers known to the survey are projected
onto a latent space consisting of the taxonomic classes from the survey 
only.\footnote{This view is slightly different from
the similarity view described in the affinity view. There,
a document includes these features but also all the rest of the 
paper metadata in terms of authors, title, abstract, venue, and so on.}

\item[-] Finally, and perhaps most interestingly, 
the above visual leads into
neighbouring (in feature space) papers that the reader 
can tap into as the state of the art 
closest to the newly generated paper(s). In addition to the metadata of
the neighbouring papers, \name surfaces the features of those 
neighbours that
need to change (either relaxed or extended) in order to make a hop from 
a known relevant paper to this non-existent paper.
\end{itemize}

\begin{figure*}
\centering
\includegraphics[width=\linewidth]{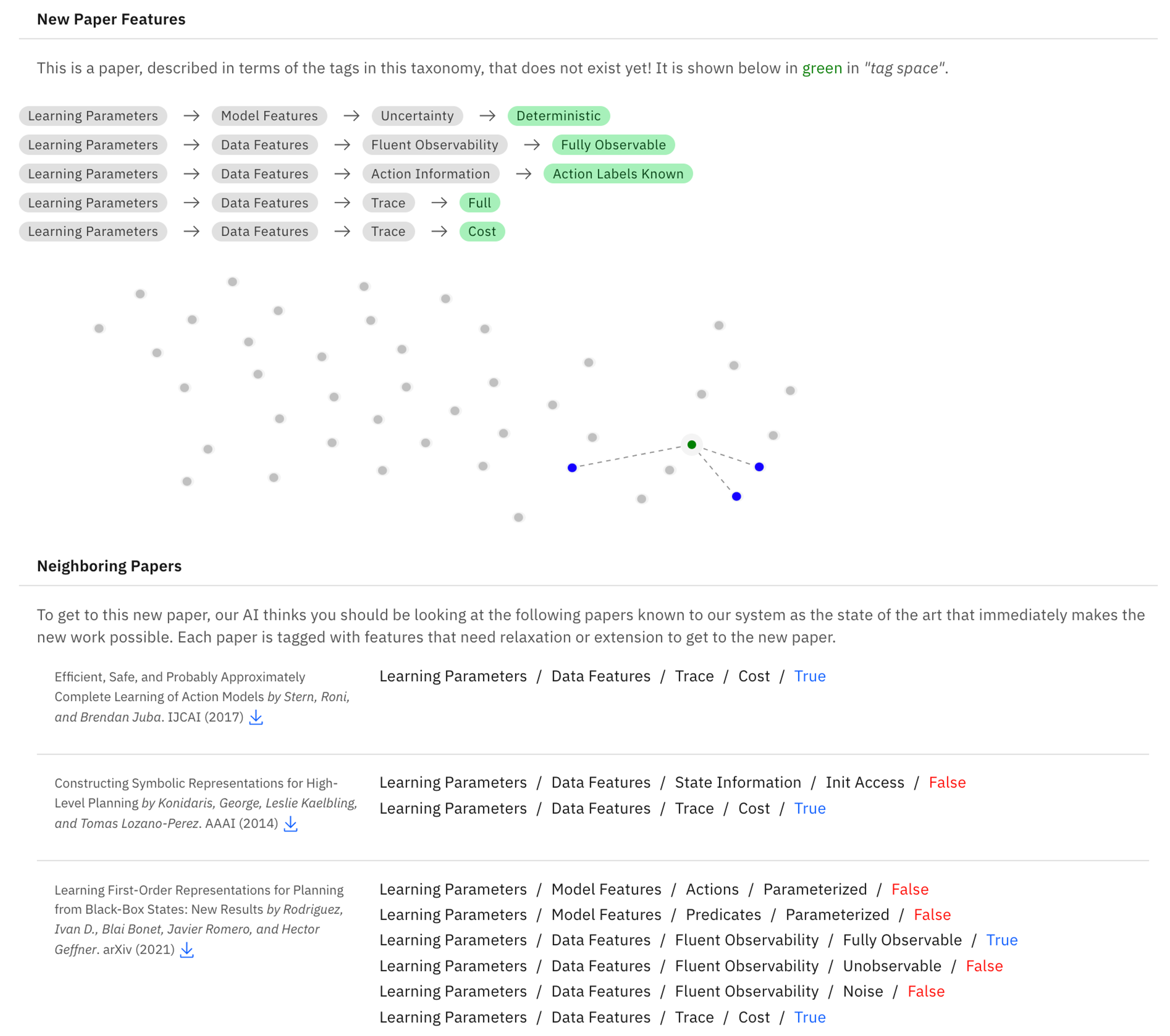}
\caption{\emph{``What paper should I write next?''} \name generating 
features of yet to be written papers
based on data from a survey, visualized in the space of tags:
the new paper is first described in terms of taxonomic classes from the underlying survey: here,
the user is recommended, in the context of model acquisition for planning, 
to look at features such as full traces with cost in the deterministic, full observable setting.
Furthermore, the system also surfaces neighboring papers known in the survey data,
whose properties or assumptions can be extended or relaxed, to arrive at this new paper.
In this instance, the system recommends that \cite{stern2017efficient} can be extended to cover traces with cost
(compare to the new properties mentioned earlier),
\cite{konidaris2014constructing} can be extended to have cover 
traces with cost as well but must be relaxed to
have no access to the initial state, and finally, the same 
for \cite{rodriguez2021learning} but by 
relaxing the assumptions on full observability of state fluents and
parameterization of actions and predicates.
For a closer look, 
we encourage the reader to try out 
this (and other use cases described later) at 
\textcolor{blue}{\href{https://macq.planning.domains}{https://macq.planning.domains}}.
}
\label{fig:teaser}
\end{figure*}

\begin{figure*}[tbp!]
\small
\begin{lstlisting}[style=yaml]
atmostone:Rationality

implies:Uncertainty > Probabilistic,Uncertainty > Non-deterministic,Uncertainty > Deterministic
atleastone:Uncertainty > Probabilistic,Uncertainty > Non-deterministic,Uncertainty > Deterministic

implies:Fluent Observability > Partially Observable,Fluent Observability > Fully Observable
implies:Fluent Observability > Unobservable,~Fluent Observability > Partially Observable
implies:Fluent Observability > Unobservable,~Fluent Observability > Fully Observable
atleastone:Fluent Observability > Unobservable,Fluent Observability > Partially Observable,
    Fluent Observability > Fully Observable

implies:Trace > Partial,Trace > Full
atleastone:Trace > Partial,Trace > Full

implies:Actions > Typing,Actions > Parameterized
implies:Predicates > Parameters Typed,Predicates > Parameterized
implies:Predicate Information > Typing,Predicate Information > Parameterized
implies:Action Information > Typing,Action Information > Parameterized
\end{lstlisting}
\caption{
Examples of custom semantic constraints defined over the taxonomic 
classes by the survey author. From deployment of: \cite{keps22}
}
\label{fig:semantics}
\end{figure*}

\section{New Paper Recommendation}
\label{sec:insights}

We now briefly describe how \name uses 
a logical theory corresponding to the \textit{valid space of research} 
according to the features contained in the survey data and manually specified constraints over them,
in order to generate yet to be written papers as a mechanism for 
exploring the research space and understanding the gaps in a field.

\subsection{Semantic Constraints over Taxonomic Classes} 

The process begins with \name auto-generating a set of relations as constraints 
to enforce from the taxonomy or classification
hierarchy contained in the survey data.
All of the papers in a survey are encoded as constraints themselves 
(their features being converted to a conjunction of Boolean variables or their negation).

This is then augmented with a set of \textit{semantic constraints} over the taxonomy,
as specified by the author of a survey -- this step is manual.
For example, in case of the deployment of \cite{keps22},
this step may describe constraints like: 
{\em ``if a technique can operate on partially observable traces, it must be able to work on fully observable traces''};
or {\em ``every technique must have full, partial, or no observability''}. 
\footnote{Interestingly, the process of specifying these constraints has one secondary benefit for the author of a survey: 
it allows one to systematically verify the documented features of the papers contained in the survey. 
This has led to several ``bug fixes'' of the data collected in \cite{keps22} already!}
Figure \ref{fig:semantics} illustrates a few such constraints on the model learning topic.

\subsection{Logical Encoding of Research Potential} 

Given the conjunction of constraints on the features, and the negation of the disjunction on the pre-existing literature (thus ruling out existing feature profiles), we have a logical theory where a satisfying assignment corresponds to a valid selection of features/assumptions about a paper, and further is one that has not yet been explored in the known literature. 
Further preferences on specific features (e.g., wanting to only handle settings without parameters for \cite{keps22}) 
can be included as unit clauses to further 
constrain the space of satisfying research configurations.
We have implemented the above encoding and integrated \name with modern 
SAT solvers \cite{kissat}
and knowledge compilers (\textsc{Dsharp} \cite{dsharp}) and a library
for representing logical sentences in negation normal form \textsc{nnf} \cite{python-nnf}), which are readily capable of handling the theory. 
Making use of a knowledge compiler and repeated logical conditioning, 
the procedure is as follows:

\begin{enumerate}
\vspace{-5pt}
\item[1.] Encode the constraints into a logical theory $T$.
\vspace{-5pt}
\item[2.] Define a full set of soft preferences $P$ over the features that stipulate ``simple'' or ``nominal'' settings (e.g., fully observable over partially observable
for \cite{keps22}).
\vspace{-5pt}
\item[3.] Run a full knowledge compilation on $T$, to get a d-DNNF representing all possible solutions $S$.
\vspace{-5pt}
\item[4.] Iterate over $p \in P$, and if $p$ is consistent with $S$, enforce $p$ by setting $S = S \wedge p$.
\end{enumerate}

\begin{table}[tbp!]
\tiny
\begin{tabular}{@{}lllclclcl@{}}
\toprule
\multirow{2}{*}{\begin{tabular}[c]{@{}l@{}}Survey\\ Instance\end{tabular}} & \multirow{2}{*}{\begin{tabular}[c]{@{}l@{}}Number \\ of Papers\end{tabular}} & \multirow{2}{*}{\begin{tabular}[c]{@{}l@{}}Number\\ of Pages\end{tabular}} & \multicolumn{2}{c}{Threshold = 0.15} & \multicolumn{2}{c}{0.25} & \multicolumn{2}{c}{0.35} \\ \cmidrule(l){4-9} 
 &  &  & \begin{tabular}[c]{@{}c@{}}Edges \\ Identified\end{tabular} & \multicolumn{1}{c}{\begin{tabular}[c]{@{}c@{}}Time \\ Taken \\ (sec)\end{tabular}} & \begin{tabular}[c]{@{}c@{}}Edges \\ Identified\end{tabular} & \multicolumn{1}{c}{\begin{tabular}[c]{@{}c@{}}Time \\ Taken \\ (sec)\end{tabular}} & \begin{tabular}[c]{@{}c@{}}Edges \\ Identified\end{tabular} & \multicolumn{1}{c}{\begin{tabular}[c]{@{}c@{}}Time \\ Taken \\ (sec)\end{tabular}} \\ \midrule
\vamhri & 175 & 1483 & 93 & 322.58 & 113 & 319.30 & 153 & 322.69\\
\macq & 43 & 934 & 84 & 138.25 & 96 & 135.72 & 110 & 134.67  \\
\xaip & 75 & 1027 & 173 & 160.72 & 173 & 158.95 & 173 & 154.11 \\
\bottomrule
\end{tabular}
\vspace{5pt}
\caption{
Automated citation network extraction performance.
}
\label{tab:network}
\end{table}

\begin{table}[tbp!]
\footnotesize
\begin{tabular}{@{}llllcll@{}}
\toprule
\multirow{2}{*}{\begin{tabular}[c]{@{}l@{}}Survey\\ Instance\end{tabular}} & \multirow{2}{*}{\begin{tabular}[c]{@{}l@{}}Number \\ of Known\\ Papers\end{tabular}} & \multirow{2}{*}{\begin{tabular}[c]{@{}l@{}}Number\\ of Tags\end{tabular}} & \multirow{2}{*}{\begin{tabular}[c]{@{}l@{}}Number of \\ papers yet to \\ be written!\end{tabular}} & \multicolumn{3}{c}{Time taken (sec)} \\ \cmidrule(l){5-7} 
 &  &  &  & k=1 & \multicolumn{1}{c}{k=2} & \multicolumn{1}{c}{k=3} \\[2ex] \midrule
\vamhri & 175 & 39 & 412,316,860,254 & 5.65 & 12.69 & 19.60 \\
\macq & 43 & 30 & 17,915,869 & 1.33 & 2.58 & 4.61 \\
\xaip & 75 & 26 & 50,331,587 & 2.36 & 4.52 & 7.11 \\ \bottomrule
\end{tabular}
\vspace{5pt}
\caption{
Automated new paper recommender performance.
}
\label{tab:imagine}
\end{table}

At the end of this procedure, we will have a single assignment to the full set of features that (1) adheres to a maximal number of preferences; and (2) differs from every other existing approach. Different orders for Step 2 will potentially result in different final outcomes.
Finally, with a candidate area of unexplored research in hand, we can perform a matching algorithm to find the closest existing approaches to the one being proposed. Our system limits this to three neighbors at a time (c.f. Figure \ref{fig:teaser}).

\section{\name in Action}

We will now briefly 
explore the three active deployments of \name,
to demonstrate the adaptability of the tool 
to different survey papers.
Brief introductions to these surveys were provided 
in Section \ref{sec:intro}.

\subsection{Parsing Performance}
\label{subsec:perf-network}

We first look at the performance of the automated citation feature 
of \name, in Table \ref{tab:network}. 
The extraction is based on segmenting each paper in a paper, 
once in single-column and again in double-column mode (whichever fits
better), and then matching phrases in the citation area to papers
known in the survey using a text similarity score.
A pre-defined match threshold indicates what fraction of a piece of text
need to be transformed to match another reference text.
Match below the threshold is taken as a citation -- this threshold is to be set 
at the discretion of the author of a survey deploying \name. 
The higher the threshold, the more connections are found but of course this will
also increase the false positive in detected citations.

\subsection{Recommender Performance}

We also look at the performance of the solver for 
the new paper recommender feature, in Table \ref{tab:imagine},
with respect to the number of papers requested by the reader.
Performance here becomes more important since this is an interactive feature. 
We also report the model count representing number of papers yet to be written
in a field as per the taxonomic classification from each survey.

\section{Conclusion}

We conclude with a brief discussion of 
a couple of research directions we are exploring at the moment.

\vspace{5pt}
\noindent {\bf Better Document Embeddings} 
The purpose of the affinity tab, as explained in Section \ref{subsec:views},
is to allow the user to visualize the papers in the space of the classification
labels introduced by the author of a survey. The visualization integrated in the 
tool currently works with basic word-based document similarity, originally used in \cite{RushStrobelt2020}, 
and does not use any features already encoded by the authors of the survey, such as the 
classification hierarchy. We are currently exploring integration with a recent
hierarchical embedding techniques \cite{yurochkin2019hierarchical} to enhance the 
view in Figure \ref{fig:dual} where papers are allocated to clusters of similar and 
neighboring concepts.

\vspace{5pt}
\noindent {\bf Interactive Recommendations} 
Finally, the ability to request features of yet unwritten papers,
is one of the most exciting features of \name. 
In order to make this more useful to researchers, we plan to perform
case studies in classes and research laboratories on the three topics -- 
model learning for planning, mixed-reality for human-robot interactions, and explainable planning -- 
explored in the surveys with deployed instances of \name, 
to determine features and use cases of value to our different use cases 
(as introduced in Section \ref{sec:intro}). 
One such feature under active development is the ability to 
select one or more taxonomic classes and / or existing papers in the database 
to focus the new paper search on, based on the interests of the user.

\vspace{5pt}
\noindent This concludes a brief overview of \name, a tool for exploring
and visualizing data from academic survey papers. 
\name is open-source at:
\textcolor{blue}{\href{https://github.com/TathagataChakraborti/survey-visualizer}{https://github.com/TathagataChakraborti/survey-visualizer}}.
We look forward to feature requests and code contributions from the 
community in the future. 

\acknowledgments{
The authors wish to thank Thao Phung (Colorado School of Mines), who was instrumental in the 
classification and analysis of the data in one of the surveys \cite{walker2022virtual} 
used in the deployment of \name.

\vspace{5pt}
\noindent A special word of appreciation for Hendrik Strobelt 
and Benjamin Hoover (MIT-IBM Watson AI Lab), 
and Sasha Rush (Cornell University, Hugging Face), whose 
works on the NeurIPS Anthology \cite{strobeltinteractive, strobeltinteractive-new}, 
and the MiniConf system \cite{RushStrobelt2020}, inspired our work.
}

\bibliographystyle{abbrv-doi-hyperref}

\bibliography{main}
\end{document}